\documentclass[10pt,journal,a4paper]{IEEEtran}
\usepackage{amsmath,amsfonts}
\usepackage{algorithmic}
\usepackage{algorithm}
\usepackage{array}
\usepackage[caption=false,font=normalsize,labelfont=sf,textfont=sf]{subfig}
\usepackage{textcomp}
\usepackage{stfloats}
\usepackage{url}
\usepackage{verbatim}
\usepackage{graphicx}
\usepackage{cite}
\usepackage{orcidlink}
\usepackage{tabularx}
\usepackage{booktabs}
\usepackage{stackengine}
\newcolumntype{L}[1]{>{\minwd l{#1}}l<{\endminwd}}
\newcolumntype{C}[1]{>{\minwd c{#1}}c<{\endminwd}}
\newcolumntype{R}[1]{>{\minwd r{#1}}r<{\endminwd}}
\def\minwd#1#2#3\endminwd{\stackengine{0pt}{#3}{\rule{#2}{0pt}}{O}{#1}{F}{F}{L}}
\newcolumntype{Q}{@{}c@{}}

\begin{document}

\title{Optimizing Large Language Models\\ for OpenAPI Code Completion}

\author{Bohdan~Petryshyn\,\orcidlink{0009-0003-4030-4842} and Mantas~Lukoševičius\,\orcidlink{0000-0001-7963-285X}\thanks{B. Petryshyn and M. Lukoševičius are with Faculty of Informatics, Kaunas University of Technology, 44249 Kaunas, Lithuania (e-mail: \{bohdan.petryshyn, mantas.lukosevicius\}@ktu.edu).}}

\maketitle

\begin{abstract}
Recent advancements in Large Language Models (LLMs) and their utilization in code generation tasks have significantly reshaped the field of software development. Despite the remarkable efficacy of code completion solutions in mainstream programming languages, their performance lags when applied to less ubiquitous formats such as OpenAPI definitions. This study evaluates the OpenAPI completion performance of GitHub Copilot, a prevalent commercial code completion tool, and proposes a set of task-specific optimizations leveraging Meta's open-source model Code Llama. A semantics-aware OpenAPI completion benchmark proposed in this research is used to perform a series of experiments through which the impact of various prompt-engineering and fine-tuning techniques on the Code Llama model's performance is analyzed. The fine-tuned Code Llama model reaches a peak correctness improvement of 55.2\% over GitHub Copilot despite utilizing 25 times fewer parameters than the commercial solution's underlying Codex model. Additionally, this research proposes an enhancement to a widely used code infilling training technique, addressing the issue of underperformance when the model is prompted with context sizes smaller than those used during training. The dataset, the benchmark, and the model fine-tuning code are made publicly available. 
\end{abstract}


\section{Introduction}
\IEEEPARstart{I}{n} the world of constantly growing digitalization and automation, Application Programming Interfaces (APIs) play a central role in enabling communication between software systems. The OpenAPI specification is a widely adopted standard for defining APIs in both machine and human-readable format \cite{openapisOpenAPISpecification}. An OpenAPI definition describes all the requests, their parameters, and responses supported by the API. Since the format is machine-readable, a variety of applications were developed to leverage OpenAPI definitions for generating API documentation and client libraries, performing automated testing, or even automatically integrating with the API using the recent advancements in Artificial Intelligence (AI) and Large Language Models (LLMs).

Although OpenAPI definitions can be fully generated based on the existing API implementation, writing a definition manually before implementing the API is a common practice that enables the development of higher-quality APIs. Writing OpenAPI definitions by hand is often a time-consuming and error-prone process due to the verbosity of the format. Modern polyglot code completion solutions like GitHub Copilot \cite{githubGitHubCopilot} and Amazon Web Services (AWS) Code Whisperer \cite{amazonCodeGenerator} have shown that LLMs can be used to increase developer productivity by providing code suggestions based on the surrounding context. These solutions are integrated into ubiquitous Integrated Development Environments (IDEs) and are capable of suggesting high-quality code completions in popular programming languages such as Python, JavaScript, and Java. However, the performance of these solutions in less popular formats like OpenAPI leaves much to be desired. 

This research evaluates the OpenAPI completion performance of one of the most widely used commercial code completion solutions, GitHub Copilot, and investigates the feasibility of developing a better-performing solution based on a modern open-source LLM Code Llama \cite{roziere2023code}. To achieve this, a semantics-aware OpenAPI completion benchmark is proposed and used to evaluate the performance of both solutions. An extensive set of experiments is performed to analyze the impact of a variety of variables, prompt engineering, and fine-tuning techniques on the performance of the Code Llama model. The results of the experiments reveal that the original version of the model, when used with optimal prompt format is already capable of outperforming GitHub Copilot in OpenAPI completion task. The fine-tuned version of the model further widens the performance gap with the commercial solution. To the best of the authors' knowledge, this is the first work investigating the Code Llama fine-tuning for code infilling in detail. The OpenAPI completion benchmark as well as the fine-tuning pipeline were made publicly available at \url{https://github.com/BohdanPetryshyn/openapi-completion-benchmark} and \url{https://github.com/BohdanPetryshyn/code-llama-fim-fine-tuning}.

\section{State of the Art in Code Generation}

As technology advances, the landscape of code generation continues to evolve, with transformer-based models and integrated solutions revolutionizing the field. This section delves into the state-of-the-art solutions for code generation, explores integrated solutions that streamline the development process, assesses methodologies for evaluating the quality of generated code, and highlights the challenges in OpenAPI autocompletion as a specific case of code generation. 

\subsection{Transformer Models for Code Generation}

Transformer-based models are the most widely used type of models for Natural Language Processing (NLP) and code generation in particular \cite{dehaerne2022code}. The original transformer model was introduced in the famous ``Attention Is All You Need'' article in 2017 \cite{vaswani2017attention}. The model was based on the trainable neural attention mechanism proposed by Bahdanau Dzmitry et al. \cite{bahdanau2014neural} earlier in 2014. This mechanism presented a solution to several problems of Recurrent Neural Network (RNN)-based models that were one of the primary methods for NLP at the time \cite{salehinejad2017recent}. 

One of the main shortcomings of RNN-based models is the difficulty of capturing long-term dependencies in the input due to vanishing and exploding gradient problems and limited hidden state size \cite{glorot2010understanding}. The attention mechanism allows the model to capture dependencies between any two tokens in the input sequence, making the likelihood of a long-term dependency being captured the same as the likelihood of any short-term dependency. 

Another major drawback of RNN-based models is the sequential nature of processing which makes it impossible to efficiently parallelize the computation. A significant part of the attention computation can be done independently for each token in the input sequence which dramatically speeds up the processing on modern highly parallel hardware like Graphics Processing Units (GPUs) and Tensor Processing Units (TPUs).

Transformer model architecture can consist of an encoder, decoder, or combination of both. Each of these parts leverages the attention mechanism in different ways to achieve its objectives. The role of an encoder is to analyze the input sequence and encode it into an internal representation that captures dependencies between the input tokens. This is done using a mechanism called self-attention where the attention score is calculated between each pair of the input tokens and a new value is calculated for each token as a sum of all the values in the input sequence weighted by the attention score. The role of a decoder is to autoregressively generate the output sequence leveraging a mechanism called encoder-decoder attention. In this process, each output token is generated as an attention-weighted sum of the encoded input tokens and the previously generated output tokens.

The encoder-only models like BERT from Google \cite{devlin2018bert}, RoBERTa \cite{liu2019roberta}, AlBERT \cite{lan2019albert}, and DeBERTa \cite{he2020deberta} only consist of an encoder and do not produce sequences as the output. Instead, these models only produce encoded representations of the input sequence that can be used for tasks like input classification, sentiment analysis, or feature extraction. The objective used during training is often not well-aligned with the inference objective, so these models usually require fine-tuning on the specific target task.

The encoder-decoder models like the original transformer suggested by Ashish Vaswani et al. \cite{vaswani2017attention}, T5 \cite{raffel2020exploring}, and BART from Meta \cite{lewis2019bart} incorporate both encoder and decoder in their architecture. These models are capable of sequence-to-sequence (seq-to-seq) generation and are commonly used for machine translation and summarization where the input and the expected output sequences can be clearly defined.

The decoder-only models like the Generative Pre-trained Transformer (GPT) family of models from OpenAI \cite{radford2018improving}, PaLM from Google \cite{chowdhery2023palm}, Chinchilla from Deepmind \cite{hoffmann2022training}, and Llama from Meta \cite{touvron2023llama, touvron2023llama2} only consist of a decoder. This preserves the ability to analyze the input sequence and generate a sequence as the output but also eliminates the boundary between the input and the output. The latter characteristic makes this type of models more generic and often abolishes the need for fine-tuning after the initial training, as most of the tasks can be represented as a natural language seq-to-seq function \cite{wang2022language}. Compared to the encoder-decoder architectures, which are also capable of seq-to-seq generation, decoder-only architectures further streamline the training process by removing the need to define the input and output formats for the model to operate on. These models are often used for language modeling and text generation but are also capable of input classification, sentiment analysis, and machine translation if invoked with a corresponding prompt. The capabilities of the large decoder-only models can be further expanded with prompting techniques like few-shot prompting \cite{reynolds2021prompt}.

Code generation is a broad term that is used to describe a wide range of activities. One such activity is code solution generation for a natural language problem definition. This task is usually approached as a seq-to-seq generation problem where the task definition is the input, and the code solution is the output of the model. One of the successful solutions in this area is AlphaCode published by DeepMind \cite{li2022competition}. The model leverages the encoder-decoder architecture. It is pre-trained on a variety of public GitHub repositories in different programming languages and fine-tuned on a curated dataset of competitive programming tasks. As stated by the authors, the model achieved a ranking of top 54.3\% among more than 5\,000 participants on the Codeforces platform.

Code generation from a natural language task definition might be useful for no-code or low-code engineers but code completion, where the model generates continuation for an incomplete code input, represents everyday programmers’ work much closer. Models like Codex from OpenAI \cite{chen2021evaluating}, CodeGen from Salesforce \cite{nijkamp2022codegen, nijkamp2023codegen2}, and Code Llama from Meta \cite{roziere2023code} leverage decoder-only architectures for autoregressive code modeling. The fill-in-the-middle (FIM) also referred to as infilling transformation and training objective allows decoder-only models to include the context from both sides of the generated part, removing the limitation of strict left-to-right generation while maintaining the ability to output long sequences of tokens \cite{bavarian2022efficient}. This approach not only enables context-aware code completion but also supports natural language task definition to some degree. These models can be prompted with a natural language code comment to generate an implementation. This can be explained by a weak natural language signal that might be captured during training on large corpora of code with comments and code-related discussions \cite{nijkamp2022codegen}.

\subsection{Transformer Model Performance Optimization}

Prompt engineering and model fine-tuning are two primary methods for large language model performance optimization. A common workflow for improving the performance of a transformer-based model in a specific domain is to use prompt engineering methods to achieve the best possible performance with the base model and then fine-tune the model on a task-specific dataset if further performance improvement is desirable. This research will utilize both prompt engineering and fine-tuning to improve the performance of the Code Llama model on the OpenAPI definitions generation task.

Prompt engineering is based on the idea that the model can be guided to generate more accurate output by optimizing the prompt in a series of human-insight-driven experiments \cite{gao2023prompt}. Common prompt engineering techniques for instruction-tuned models include few-shot prompting \cite{reynolds2021prompt}, chain-of-thought prompting \cite{wei2022chain}, and generated knowledge prompting \cite{liu2021generated}. For autoregressive code generation, prompt engineering techniques include embedding additional context to the prompt in the form of comments or other language-specific constructs \cite{nijkamp2022codegen} and optimizing the prompt structure and size to maximize the model's performance and resource efficiency.

Fine-tuning is the second primary method for improving the performance of large language models. The idea behind this method is that the model that has already been trained for a similar or broader task can be further fine-tuned for a specific task. The fine-tuning process is similar to training, but it is performed on a pre-trained model and requires significantly less data and computational resources. Two fine-tuning approaches are commonly used: full fine-tuning and Parameter-Efficient Fine-Tuning (PEFT) \cite{ding2023parameter}.

Full fine-tuning is performed by continuing the training process of an already pre-trained model. During this process, all the model parameters are updated just like during the original training. This means that the hardware requirements for the full fine-tuning are also similar. This approach demonstrates the potential for achieving the best results, but it also requires the most computational resources and data. The ability to change all the model parameters during full fine-tuning can also lead to catastrophic forgetting, a training failure mode where the model loses the ability to perform the original task or stops generating adequate outputs altogether.

Parameter-efficient fine-tuning, as an alternative to full fine-tuning, is performed by freezing most of the pre-trained model parameters and only training a relatively small number of additional parameters. This approach requires much less computational resources and tends to yield comparable results to the full-finetuning \cite{ding2023parameter}. The PEFT approach is also considered to be less prone to catastrophic forgetting as the majority of the pre-trained model parameters are left unchanged during the fine-tuning process. Notable PEFT techniques include adapter tuning \cite{houlsby2019parameter} which inserts additional adapter layers between the original model layers and only updates them during training, BitFit \cite{zaken2021bitfit} which only trains the bias-terms of the original model, and Low-Rank Adaptation (LoRA) \cite{hu2021lora} which decomposes the original parameters into low-rank matrices and only trains them. The LoRA method claims to reduce the number of trainable parameters by 10\,000 times and memory requirements by 3 times while maintaining or surpassing the performance of full fine-tuning when applied to models like GPT-3 and RoBERTa.

\subsection{Integrated Code Generation Solutions}

Integrated code generation and completion solutions are end-to-end systems that combine code generation models, prompt engineering techniques, developer experience optimizations, and IDE integrations to solve a real-world problem of improving developer productivity. These solutions are often proprietary, so the details of their implementation are not available to the public. However, some solutions are open-source and can be analyzed in detail.

The most notable commercial solution is GitHub Copilot. Developed by GitHub, OpenAI, and Microsoft, this code assistant is based on the Codex \cite{chen2021evaluating} model. The solution is integrated into several IDEs like Visual Studio Code, NeoVim, and JetBrains IDEs. As discovered by a reverse engineering effort \cite{thakkarparth007Copilotexplorer}, Copilot's prompt-building approach is based on several language-specific heuristics. For example, for programming languages like Python, a list of imported libraries and declared functions within the file is extracted and always included in the prompt. The code assistant underwent manual evaluation in several studies \cite{nguyen2022empirical, sobania2022choose, dakhel2023github, wermelinger2023using} demonstrating decent performance in benchmarks like HumanEval \cite{chen2021evaluating} and solving moderate complexity tasks at platforms like LeetCode. No studies were found that would evaluate the solution performance in OpenAPI completion. In this research, GitHub Copilot is evaluated as the baseline solution for the OpenAPI definitions completion task.

Other proprietary solutions worth mentioning are Tabnine \cite{tabnineTabnineAssistant} which emphasizes the security of the user's data and the code base, AWS Code Whisperer \cite{amazonCodeGenerator} which demonstrates additional expertise in AWS services and client libraries and performs additional checks for license compliance and common security vulnerabilities in the generated code, and Sourcegraph Cody \cite{sourcegraphCodyCoding} which demonstrates awareness of the whole user's code base.

Continue \cite{githubGitHubContinuedevcontinue} is an example of an open-source integrated code completion solution that aims to be model-agnostic and supports multiple models with the most commonly used being Code Llama Instruct running in a local or a cloud-based LLM server. Other open-source models supported by the solution include WizardCoder \cite{luo2023wizardcoder}, Phind Code Llama \cite{phindmodel}, Mistral \cite{jiang2023mistral}, CodeUp \cite{githubGitHubJuyongjiangCodeUp}, and Zephyr \cite{tunstall2023zephyr}. Commercial models like GPT-3, GPT-4, and PaLM-2 \cite{anil2023palm} can be integrated as well. During code assessment, it was discovered that Continue prompts the model in an instructional manner by providing it with the surrounding code and asking to generate the missing part. This explains the fact that the solution only works with models that support instruction execution. Another open-source solution, LLM-vscode \cite{githubGitHubHuggingfacellmvscode}, is a Visual Studio Code extension developed by Hugging Face for demonstrational purposes. The extension can be used with many models but in contrast to Continue, it supports both instruction execution and code infilling models. LLM-vscode is usually used with models like Code Llama, Phind Code Llama, and WizardCoder.

Another research on applying LLMs to a real-world code completion use case was published by Microsoft in 2020 \cite{svyatkovskiy2020intellicode}. Apart from evaluating transformer optimization techniques for code like applying tokenization methods that can overcome the closed vocabulary problem, this research describes multiple developer experience optimization techniques like client-side tree-based caching, parallel beam search decoder, and compute graph optimizations.

\subsection{Code Generation Evaluation}

Evaluation of LLMs has always been a challenging task in the field of NLP. The evaluation methods have been evolving in step with the abilities of the models and the complexity of the tasks they perform.

As the first applications of language models started appearing in the field of natural language modeling and machine translation, the evaluation methods were designed to measure the quality of the generated text by comparing it to the expected output. The most widely used metrics for this purpose are BLEU \cite{papineni2002bleu}, ROUGE \cite{lin2004rouge}, and METEOR \cite{banerjee2005meteor}. These metrics are based on calculating the similarity by counting the overlapping n-grams or measuring the longest common sequences. 

Code generation can not be reliably evaluated with machine translation metrics due to the fundamental differences between natural language and code. The code is structured and possesses a more rigid syntax than natural language, making it more sensitive to small character-level changes. Another source of complexity in code evaluation is the fact that a single code problem can have multiple equally correct solutions. This makes it impossible to define an exhaustive set of expected outputs for a given input. 

One of the current state-of-the-art evaluation methods for code generation is the HumanEval benchmark \cite{chen2021evaluating}. Originally proposed by OpenAI, this benchmark was designed to evaluate Python code generation by providing a set of problems each with a description and a set of automated test cases. Later, the method was extended to support 18 additional programming languages by the MultiPL-E benchmark \cite{cassano2023multipl}. These methods take language semantics into account but are only applicable to a limited set of programming languages and do not support specific code semantics like OpenAPI definitions.

The OpenAPI specification is not a programming language but a data format specification. This means that the existing code generation evaluation methods can not be easily extended to support OpenAPI completion evaluation. To enable the development of OpenAPI completion solutions, a new benchmark has to be proposed that supports the specific semantics of OpenAPI definitions.

\section{Methods and Datasets}

As outlined in the previous section, OpenAPI completion requires specialized evaluation and optimization methods. This section describes the proposed OpenAPI completion benchmark, the foundational model selected for the study as well as the prompt engineering and fine-tuning techniques used to optimize the model's performance.

\subsection{OpenAPI Completion Evaluation Criteria and Benchmarking}

As discussed in the previous section, existing text or code generation evaluation methods can not be considered optimal for OpenAPI generation evaluation. These methods are not semantics-aware as in the case with machine language translation metrics or do not support specific OpenAPI semantics as in the case with the widely used HumanEval and MultiPL-E benchmarks.

\begin{figure}[!h]
\centering
\includegraphics[width=\columnwidth]{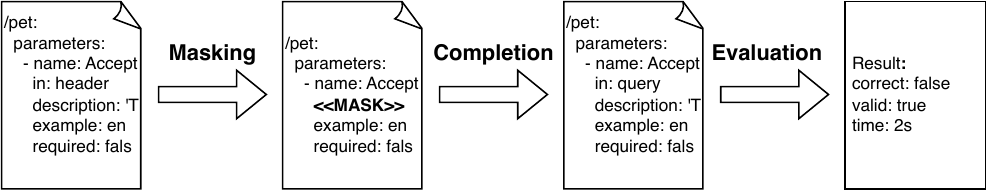}
\caption{Proposed OpenAPI completion benchmark pipeline.\label{fig:benchmark-pipeline}}
\end{figure}

The proposed benchmark aims to closely resemble the real-world usage of the solution in terms of input modeling and output evaluation. Figure~\ref{fig:benchmark-pipeline} outlines the steps for a single evaluation iteration. The masking step, where part of the input definition is replaced with a special mask token, is responsible for modeling the input. To equalize the evaluation conditions for all the solutions, the same set of masked test cases is used for all the evaluations. At the completion step, the evaluated model is used to replace the mask token with the generated OpenAPI definition. The evaluation step, where the completed definition is compared to the original, is responsible for the completion result evaluation.

The masking step can be seen as splitting the original OpenAPI definition into two parts (prefix and suffix) and removing some amount of the definition from in between. This way, the input modeling is defined by the position of the prefix-suffix split and the size of the masked part between them. When editing an OpenAPI definition, the user does not always add new endpoints or parameters and can expect to get a completion suggestion even when they have not finished typing the previous word or syntactic construct. This means that the prefix and suffix split should be randomly selected and not tied to a newline or a keyword. Conversely, the end of the masked sequence should be on a newline, as it is reasonable to assume that users are more likely to edit a document at the end of a line rather than in the middle of it when describing new API operations or adding parameters to existing ones. In this study, the masked part is always 10 lines long (till the end of the 10th line), which is a reasonable size for a single parameter or a response definition.

The following metrics are measured at the evaluation step:
\begin{enumerate}
\item Correctness -- how often the generated OpenAPI definition is semantically identical to the original one;
\item Validity -- how often the generated OpenAPI definition is syntactically valid;
\item Speed -- how fast the solution generates the OpenAPI definition.
\end{enumerate}

To be semantically identical, two snippets of OpenAPI definition do not have to be identical at the character level. The YAML and JSON specifications allow for different ways of representing the same data. For example, keys can be optionally quoted, and the order of keys is not significant. The OpenAPI specification allows for even more flexibility with the support of references and other advanced syntactic features. Additionally, OpenAPI definitions contain subjective information such as descriptions, titles, summaries, and examples of data. These fields are not required to be identical for the generated OpenAPI definition to be considered semantically identical to the expected one.

To evaluate correctness, the benchmark uses the Oasdiff tool \cite{oasdiffOpenAPIDiff} which is capable of calculating the semantic difference between two OpenAPI definitions. The calculated difference is not affected by the YAML, JSON, or OpenAPI specification flexibility described above. To tolerate subjectiveness in the definition, the following set of heuristics is applied to the calculated difference:
\begin{enumerate}
\item The definitions are semantically identical if the calculated difference is empty;
\item The definitions are semantically identical if the calculated difference only contains insignificant changes;
\item Changes in the following subjective fields are considered insignificant: description, summary, title;
\item Changes in the examples of data are considered insignificant if they do not change the structure and type of the data.
\end{enumerate}

\subsubsection{\bf Evaluation dataset}

The dataset for the benchmark was collected from the APIs-guru definitions directory \cite{githubGitHubAPIsguruasyncapidirectory} which is a weekly-updated collection of public OpenAPI definitions. The directory contains more than 4\,000 definitions in YAML format.

Ideally, the definitions have to be unseen by the model during training or fine-tuning. However, as the Code Llama model authors do not reveal the details of the training dataset, it is not possible to guarantee that the definitions were not used during training. Only the newest definitions in the directory were selected for the evaluation dataset to minimize the chance of being seen by the model. The definitions were also checked for presence in The Stack \cite{huggingfaceBigcodethestackDatasets} dataset which is one of the biggest open code datasets collected from GitHub. The definitions that were present in The Stack dataset were excluded from the evaluation dataset.

Another requirement for the definitions in the evaluation dataset is that they have to represent different domains and styles. The APIs-guru collection contains definitions for APIs from a variety of domains such as payment systems, cloud services, and electronic commerce. The directory also contains definitions from many producers such as Amazon, Meta, and Google. The definitions were selected from different domains and producers to ensure that the benchmark is not biased toward a specific field or style of OpenAPI definitions. Automatically generated definitions were detected and excluded from the evaluation dataset as they do not represent the future real-world use of the solution.

Only OpenAPI definitions larger than 3\,000 lines were selected for the evaluation dataset with the preference given to definitions larger than 20\,000 lines when all the other criteria are equal. Large definitions are preferred since they are more likely to contain complex syntactic constructs and semantic dependencies. Small definitions that fit into the context size of the model are often too trivial to complete correctly and do not present any practical value for the benchmark.

Ten OpenAPI definitions satisfying the above criteria were selected for the evaluation dataset (Table~\ref{tab:benchmark-definitions}). Considering the size of the selected definitions, multiple test cases could be extracted from each definition. The test cases were extracted by randomly selecting a position in the definition and masking the rest of the line and ten following lines. The evaluation objective was to generate the infilling for the masked part so that the resulting definition is semantically identical to the original one.

\begin{table}[!h]
\caption{OpenAPI Definitions Included in the Evaluation Dataset\label{tab:benchmark-definitions}}
\centering
\begin{tabular}{lllr}
\toprule
\textbf{API Title} & \textbf{Producer} & \textbf{Doman} & \textbf{Size, lines} \\
\midrule
Adyen API & Adyen & Payments & 20\,359 \\
Google Video API & Google & Video & 16\,008 \\
Mux API & Mux & Video & 7\,128 \\
Rubrik API & Rubrik & Security & 43\,540 \\
Severa Public API & Visma & Project mngmt. & 36\,871 \\
Sinao API & Sinao & Accounting & 12\,215 \\
SpaceTraders API & SpaceTraders & Gaming & 3\,121 \\
The Racing API & The Racing API & Sports & 9\,650 \\
VPC Lattice API & AWS & Cloud & 6\,709 \\
Visma e-conomic & Visma & Accounting & 10\,162 \\
\bottomrule
\end{tabular}
\end{table}

The benchmark was implemented with the idea of being fully automated and reproducible. This made it possible to easily evaluate the performance of many solutions and configurations. However, the fact that commercial solutions like GitHub Copilot are only available as IDE plugins and require manual interaction with the IDE imposed a limitation on the size of the evaluation dataset. To keep the manual evaluation time reasonable while still providing a representative evaluation, the dataset size was constrained to 100 test cases in total (10 test cases from each of the 10 definitions) which is comparable to the size of the HumanEval benchmark with 164 code problems.

\subsection{Code Llama Model}\label{sec:code-llama}

Training LLMs from scratch is a highly computationally expensive process \cite{hoffmann2022training}. For practical reasons, it is often more efficient to incorporate a generic pre-trained model that can be fine-tuned for a specific task or used in a zero-shot manner. For this research, the Code Llama family of models from Meta \cite{roziere2023code} was selected as the foundational model for the OpenAPI completion task.

The Code Llama model is trained on an unbound set of programming languages which not only ensures its ability to generate OpenAPI definitions but also allows it to generate code samples that are commonly found in OpenAPI definitions in a variety of programming languages. The model supports large context sizes (up to 100\,000 tokens) which is important for the OpenAPI completion task as the definitions can reach tens of thousands of lines in size.

\begin{figure}[!h]
\centering
\includegraphics[width=\columnwidth]{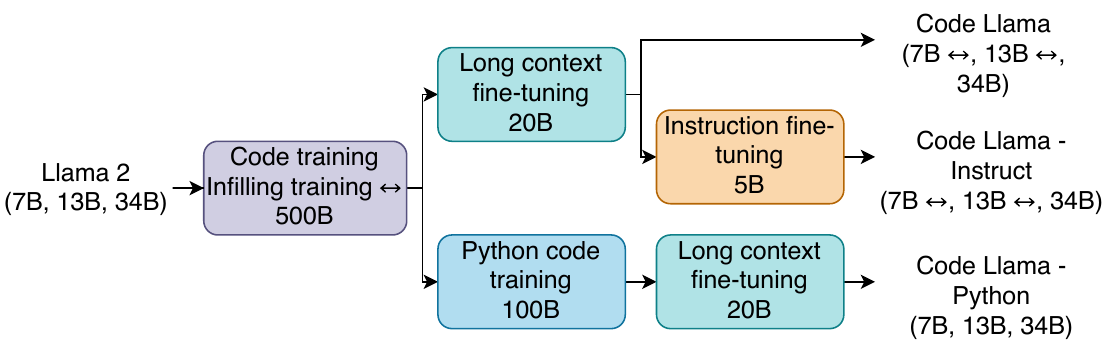}
\caption{Code Llama model family training pipeline, adapted from \cite{roziere2023code}.\label{fig:code-llama-family}}
\end{figure}

This code generation model is based on another widely adopted natural language model from the same company -- Llama 2. The model was trained for code infilling on an additional 500 billion tokens of code. Figure~\ref{fig:code-llama-family} outlines the training pipeline of the Code Llama model family. The model was further trained for Python specialization and fine-tuned for instruction execution, but for the sake of this research, the base Code Llama model is used. The 70-billion-parameter version of the model was trained and released months after the original Code Llama family and might follow a slightly different training pipeline.

After the recent release of the biggest version, the model is now available in four sizes: 7, 13, 34, and 70 billion parameters. This study is focused on the smaller 7 and 13-billion-parameter versions of the model because the bigger 34-billion-parameter version does not support the infilling mode of operation (``$\leftrightarrow$'' in Figure~\ref{fig:code-llama-family}, discussed below) and the computational requirements of the 70-billion-parameter version exceed the capabilities of the evaluation platform.

The Code Llama model can operate in two modes: autoregressive code generation, where the model takes a code snippet and generates a continuation for it token by token, and code infilling, where the model takes two snippets of code and generates the completion that would fill the gap between them. The mode of operation is determined by the format of the prompt. For code generation, the model should be prompted with the code snippet as is. For code infilling, the two snippets (prefix and suffix) should be formatted in a special way that was defined during the model training.

Following the training methods originally described in the work of Bavarian et al. \cite{bavarian2022efficient}, the Code Llama model supports two prompt formats for code infilling: Prefix Suffix Middle (PSM) and Suffix Prefix Middle (SPM). The study has shown that the SPM format produces slightly better results even when the model is trained on the PSM format. The SPM format is also expected to result in a higher key-value cache hit rate as the suffix is less likely to change in a real-world code completion scenario. Both formats are evaluated in this research.

The Code Llama model introduces four special tokens to the existing Llama 2 tokenizer: $<$PRE$>$, $<$SUF$>$, $<$MID$>$, $<$EOT$>$. Using these tokens, the PSM prompt format is defined in \cite{bavarian2022efficient} as 
\begin{multline}
\text{$<$PRE$>$} \circ \text{Enc(prefix)} \circ \text{$<$SUF$>$}\circ \\
\text{Enc(suffix)} \circ \text{$<$MID$>$} \circ \text{Enc(middle)},
\label{eq:PSM}
\end{multline}
where the Enc$(\cdot)$ is the encoding operation and $\circ$ is a token-level concatenation. The SPM format is defined in \cite{bavarian2022efficient} as
\begin{multline}
\text{$<$PRE$>$} \circ \text{$<$SUF$>$} \circ \text{Enc(suffix)} \circ \\
\text{$<$MID$>$} \circ \text{Enc(prefix)} \circ \text{Enc(middle)}.
\label{eq:SPM}
\end{multline}
The model autoregressively generates tokens until the $<$EOT$>$ token is generated indicating a successful connection of the middle part with the suffix. The fact that in the SPM format, the middle part immediately follows the prefix makes the format even more similar to the simple regression case.

\subsection{Prompt Engineering Techniques for OpenAPI Completion}\label{sec:prompt-engineering-methods}

As the expected OpenAPI definition size is significantly larger than the maximum context size supported by Code Llama, it is not possible to fit the whole definition in the prompt. In this research, the impact of multiple variables on the completion performance is evaluated to find the optimal prompt format for a real-world OpenAPI completion task. 

Code Llama's infilling mode is more suitable for the OpenAPI completion task than the autoregressive generation mode as it allows the model to consider the context both before and after the cursor position. The part of the definition that goes after the cursor should allow the model to take into account components such as schemas, requests, and responses that are commonly placed at the end of the document. The suffix is also important for the model to generate syntactically correct definitions that are integral to the rest of the document. Editing at the end of a document can be considered an infilling task where the suffix is empty. In the real world, API designers are more likely to edit existing definitions by adding new operations or parameters in the middle of the document rather than at the end of it.

One of the factors affecting completion performance is the prefix-to-suffix ratio. Theoretically, prefixes and suffixes should contain the same amount of useful information for the model to generate a semantically correct completion. However, the model can have a preference for one of the parts inherited from the training data. The optimal ratio can be found by evaluating the completion performance for a range of ratios. 

Another factor that affects both the completion accuracy and speed is the context size. The context size is the number of tokens available to the model at the time of completion. A larger context size allows the model to take into consideration more information from the surrounding context but also requires more computational time and resources to process. The optimal context size can be found by evaluating the completion performance for a range of context sizes while keeping the prefix-to-suffix ratio constant. 

OpenAPI specification allows including metadata about the API in the definition. This metadata can include information like the API's title, description, version, contact information, or license type. The metadata is not required for the definition to be considered valid and can be neglected during OpenAPI validation or comparison. However, the metadata may contain useful information for the model to generate a more semantically correct definition.

The components section of the OpenAPI definition is crucial for the model to generate valid definitions since it defines the set of available schemas, requests, responses, and other components that can be referenced in the definition. To test this hypothesis, the prompt builder is modified to extract the components section if available, format it as a comment with the list of component names, and include it in the prompt prefix.

\subsection{Code Llama Model Fine-tuning for OpenAPI Completion}\label{sec:fine-tuning-methods}

The OpenAPI completion performance of the model can be further improved by fine-tuning it on task-specific data. This section describes the dataset used for fine-tuning the 7-billion-parameter version of the Code Llama model as well as the full details of the fine-tuning process.

The dataset for fine-tuning was collected from the APIs-guru OpenAPI definitions directory \cite{githubGitHubAPIsguruasyncapidirectory}. The directory contains more than 4\,000 definitions in YAML format. Analysis of the repository revealed that about 75\% of the definitions in the directory are produced by a handful of major companies like Amazon, Google, and Microsoft. To avoid the dataset bias towards a specific producer, the maximum number of definitions from a single producer was limited to 20. Multiple versions of the same API were also excluded from the dataset as they are likely to contain similar definitions. Definitions from producers used in the evaluation dataset were also excluded from the fine-tuning dataset to avoid data leakage. The resulting dataset contains 990 definitions and is accessible at the HuggingFace Hub at \url{https://huggingface.co/datasets/BohdanPetryshyn/openapi-completion-refined}.

To avoid the objective mismatch between the training and fine-tuning processes, the infilling prompt format was reconstructed to mirror the format described by the Code Llama model authors. Nevertheless, the specifics of the OpenAPI format had to be taken into account. Similarly to the Code Llama training, the prefix-middle-suffix split was performed at the character level with the split positions selected randomly. The same tokens ($<$PRE$>$, $<$SUF$>$, $<$MID$>$, $<$EOT$>$) were used to format the prompt in the PSM \eqref{eq:PSM} and SPM \eqref{eq:SPM} formats and the leading spaces implicitly generated by the SentencePiece tokenizer were suppressed. However, unlike the Code Llama training, the split was performed at the context level and was not limited to documents that fit into the context size as most of the OpenAPI definitions are much larger than the context size used during fine-tuning. In the SPM format, the prefix and middle parts were concatenated after the tokenization to improve the model performance with split tokens which are expected to be encountered often in the real-world OpenAPI completion scenarios.

\begin{figure}[!h]
\centering
\includegraphics[width=\columnwidth]{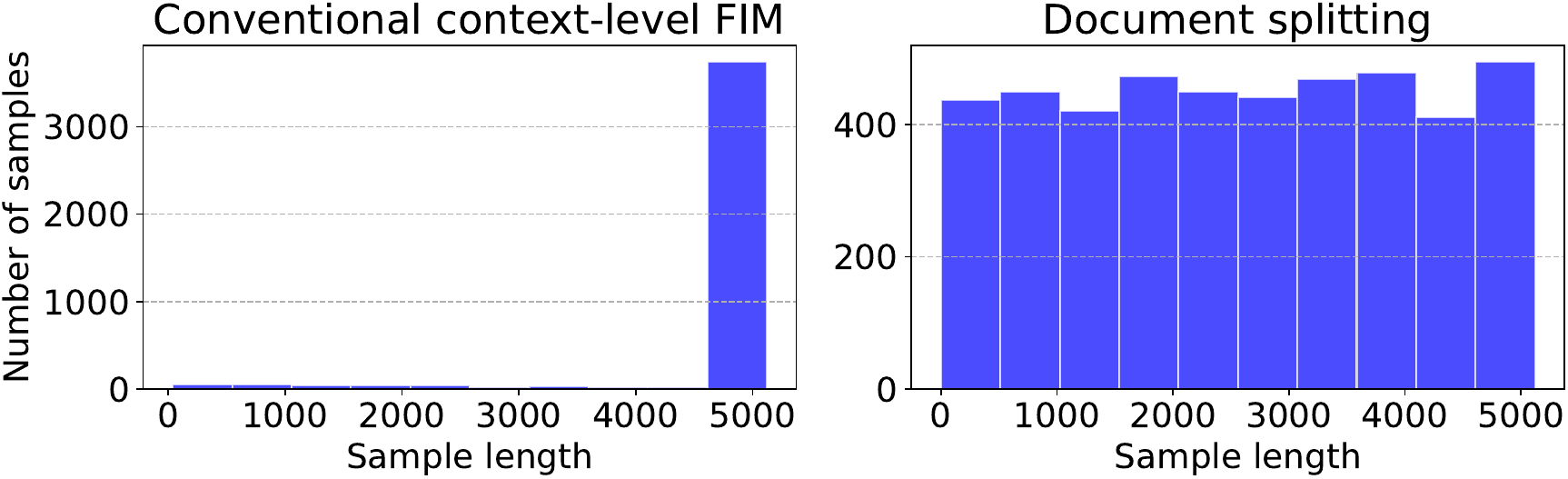}
\caption{Training sample length distribution with and without the document splitting technique.\label{fig:sample-length-distribution}}
\end{figure}

Another consequence of the large size of the OpenAPI definitions is the bias towards training-time context size infilling that might be developed when fine-tuning the model using the conventional FIM training objective as described in the work of Bavarian et. al. \cite{bavarian2022efficient}. This bias can cause the model to show optimal performance only when prompted with the same context size as during fine-tuning. As can be seen in Figure~\ref{fig:sample-length-distribution}, the distribution of the training sample lengths is skewed toward the maximum context size of 5\,120 tokens when the conventional context-level FIM transformation is used. To mitigate the bias, the input documents were split into smaller pieces of random length between 4\,096 and 6\,144 tokens (0.8-1.2 of the 5\,120 tokens context size) before proceeding to the context-level FIM transformation. This made sure the context chunks included multiple documents of various lengths more often and evened the distribution of the training sample lengths. This technique is further referred to as document splitting. When used before the FIM transformation, document splitting should not be distinguishable from using full documents of uniform length distribution as the truncated prefixes and suffixes correspond to a normal real-world code completion scenario when the context size is limited and does not cover the entire document.

Parameter efficient fine-tuning was performed with the following hyperparameters: LoRA rank of 32, LoRA alpha of 64, LoRA dropout of 0.1, AdamW optimizer with the learning rate of 2e-4, cosine scheduler with the warmup ratio of 0.1, the batch size of 4 with the gradient accumulation steps of 8 and the maximum number of optimization steps of 1\,000.

\section{Results}

Following the prompt engineering and fine-tuning methods described in the previous section, the Code Llama model was optimized for the OpenAPI completion task. This section presents the results of the evaluation of the optimized model and compares it to the performance of GitHub Copilot. For all the evaluations described in this section, the same set of masked OpenAPI definitions was used ensuring consistent evaluation conditions.

\subsection{GitHub Copilot Performance in OpenAPI Completion}

GitHub Copilot was selected as the baseline solution for OpenAPI generation as at the time of writing it is the most widely used coding assistant on the market. The solution is a commercial product available with a monthly subscription plan.

As GitHub Copilot does not provide an API for programmatic access, the completion cannot be automated and must be performed manually. For the evaluation results to be comparable with the automatic completion results, the manual evaluation process resembles the automatic completion procedure as closely as possible and uses the same set of masked OpenAPI definitions. Manual completion was performed for each masked definition following these steps:

\begin{enumerate}
\item The $<<$MASK$>>$ marker is removed, and the cursor is set in place of the marker;
\item If Copilot refuses to generate a suggestion, the last character is removed and added again to trigger the completion;
\item Copilot’s suggestions are accepted without modifications one by one until one of the following conditions is reached:
\begin{enumerate}
\item Suggestions are no longer generated (Copilot considers the infilling complete);
\item Maximum number of completion lines is reached (15 lines).
\end{enumerate}
\end{enumerate}

The speed of suggestion generation using GitHub Copilot was measured by tracking the time between triggering completion with a key press and seeing the generated suggestion on the screen. The number of characters generated was divided by the generation time. The experiment was repeated ten times to take human error as well as random service latency factors into account. The average code generation speed was measured to be 19.3 characters per second with a standard deviation of 1.9.

Following the completion procedure described, GitHub Copilot correctly completed 29\% of the OpenAPI definitions. 68\% of the completed definitions were valid.

\subsection{Original Code Llama Model for OpenAPI Completion}

Using the benchmark proposed in the previous chapter, an extensive set of experiments was performed to analyze the impact of a variety of variables and prompt-engineering techniques on the performance of the Code Llama model in the OpenAPI completion task. The experiments were designed to analyze the impact of a single variable at a time. When not stated otherwise, the 7-billion-parameter version of the Code Llama model, the PSM prompt format \eqref{eq:PSM}, the 50/50 prefix-to-suffix ratio, and the 4\,096 tokens context size were used.

The Code Llama models were evaluated at the Hugging Face Inference Endpoints platform \cite{huggingfaceInferenceEndpoints} using the simple greedy search decoding strategy. Without quantization, a single Nvidia A10G GPU was used for the 7-billion-parameter model. For the 13-billion-parameter model, four Nvidia A10G GPUs were used.

\subsubsection{\bf Optimal prefix-to-suffix ratio}

A range of prefix-to-suffix ratios from 30/70 up to 90/10 was evaluated to find the optimal proportion. The evaluation was performed on the 7-billion-parameter version of the Code Llama model with a constant context size of 4\,096 tokens. The optimal prefix-to-suffix ratio can depend on the above-mentioned factors but for parameter space reduction, the dependencies are intentionally disregarded in this research.

\begin{figure}[!h]
\centering
\includegraphics[width=\columnwidth]{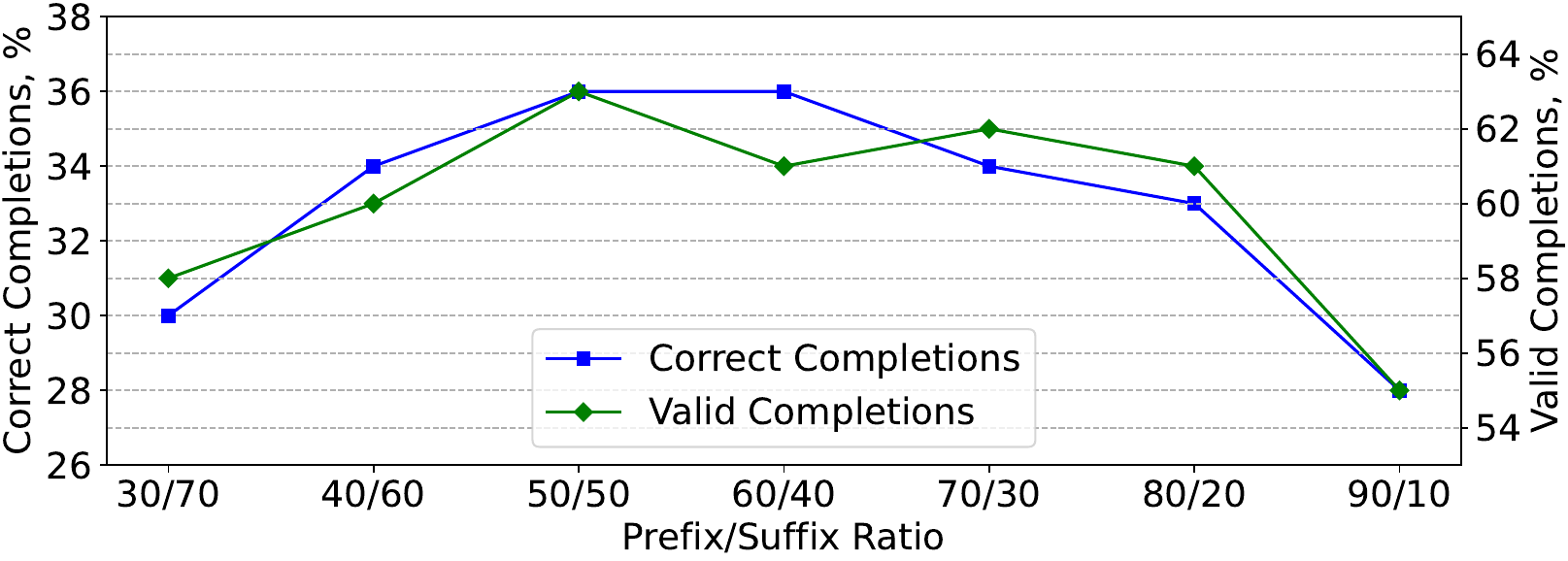}
\caption{OpenAPI completion performance of the original Code Llama 7B model for different prefix-to-suffix ratios with a fixed context size of 4\,096 tokens.}
\label{fig:prefix-to-suffix-ratio}
\end{figure}

Figure~\ref{fig:prefix-to-suffix-ratio} depicts the dependency of correctness and validity of the OpenAPI completion on the prefix-to-suffix ratio. As can be seen from the graph, the highest percentage of correctly infilled definitions of 36\% corresponds to 50/50 and 60/40 prefix-to-suffix ratios. The 50/50 ratio demonstrates a higher validity percentage of 63\% compared to 61\% for the 60/40 ratio. Even though the dip at the 60/40 ratio can be attributed to the factor of randomness, the 50/50 ratio can still be considered optimal as it does not discriminate the importance of either prefix or suffix without sound justification.

\subsubsection{\bf Optimal context size}

Context size is another factor that affects not only the solution performance but also the speed of code generation and, consequently, throughput and resource efficiency of a real-world implementation. A range of context sizes from 1\,024 to 7\,168 tokens was evaluated with a constant prefix-to-suffix ratio of 50/50 and the 7-billion-parameter version of the Code Llama model.

\begin{figure}[!h]
\centering
\includegraphics[width=\columnwidth]{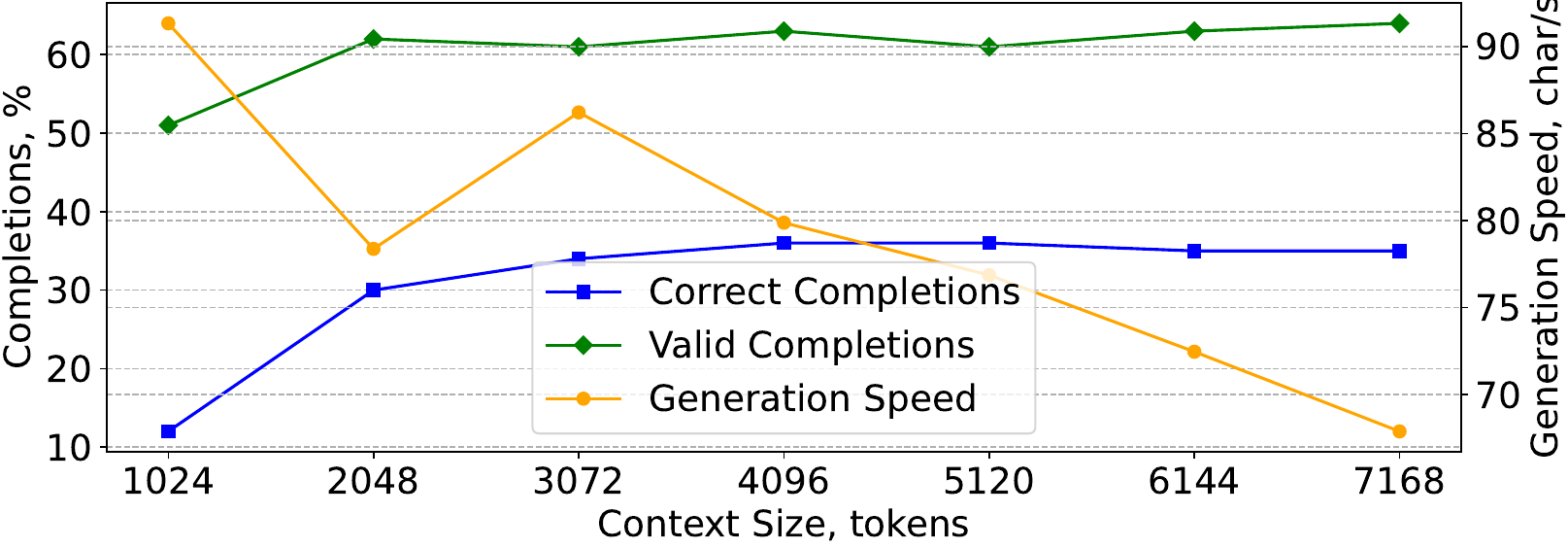}
\caption{OpenAPI completion performance of the original Code Llama 7B model for different context sizes with a fixed preffix-to-suffix-ratio of 50/50.\label{fig:context-size}}
\end{figure}

As can be seen in Figure~\ref{fig:context-size}, OpenAPI completion correctness and validity increase with the expanding context size. The correctness rate, which is the primary target metric in this research, demonstrates a much smoother trend with a significant increase up to the context size of 4\,096 tokens. After this point, the correctness level stabilizes at around 36\% and even decreases by 1\% after 6\,144 tokens. The speed, as expected, follows the downward trend with a dramatic decline after the 3\,072 tokens point. This can be explained by the quadratic complexity of the input processing in decoder-only transformers.

With a correctness rate of 36\%, validity rate of 63\%, and a generation speed of 79 characters per second, the context size of 4\,096 tokens can be considered optimal for real-world scenarios since after this point, neglectable correctness and validity increases come at a cost of significant speed and resource-efficiency deterioration.

\subsubsection{\bf PSM and SPM infilling formats}\label{sec:spm-vs-psm}

\begin{figure}[!h]
\centering
\includegraphics[width=\columnwidth]{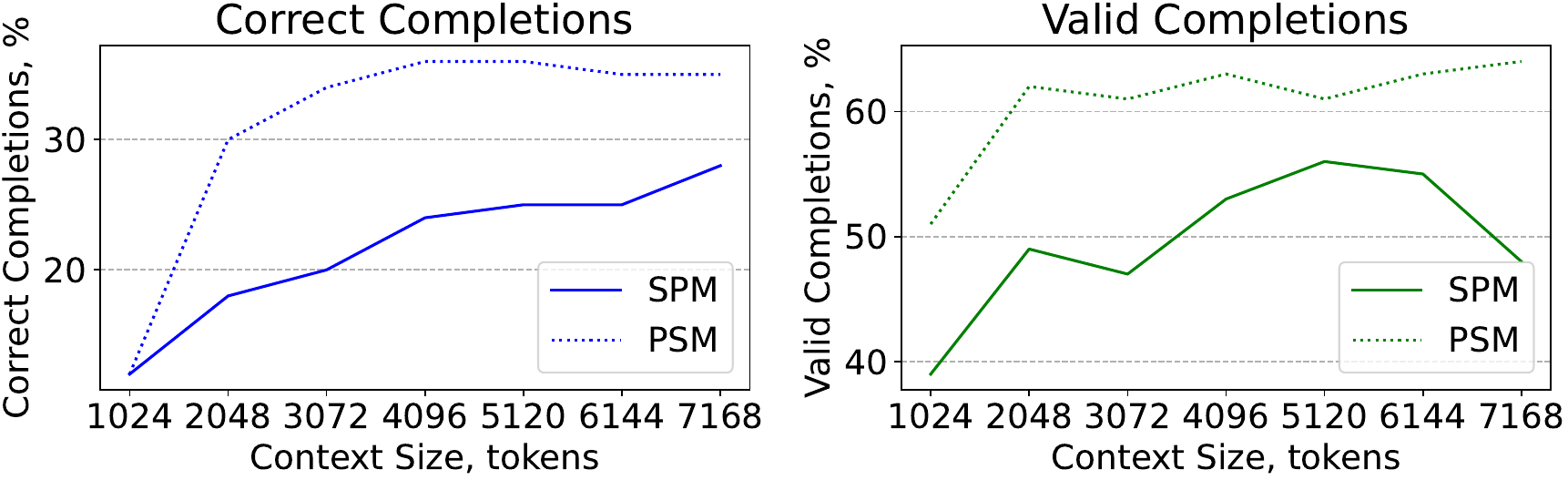}
\caption{OpenAPI completion performance of the original Code Llama 7B model with the PSM and SPM prompt formats.\label{fig:spm-vs-psm}}
\end{figure}

SPM format \eqref{eq:SPM} is an alternative prompt format supported by the Code Llama model in the infilling mode. Contrary to the fact that the SPM format demonstrated higher performance in the original work of Bavarian et al. \cite{bavarian2022efficient}, the authors of Code Llama designed PSM \eqref{eq:PSM} to be the primary prompting format for the model stating that the SPM implementation might require additional token healing on the tokenization step at the inference time. In this experiment, the default Code Llama tokenizer is used. As can be seen in Figure~\ref{fig:spm-vs-psm}, the SPM prompting format underperforms at all context sizes compared to PSM. For example, the model correctly completed 24\% of the definitions and generated valid completions 53\% of the time at 4\,096 tokens, which is a 33.3\% correctness decrease compared to the same configuration with the PSM prompt format.

\subsubsection{\bf OpenAPI metadata in prompt}

This experiment evaluates the impact of including the components section of the OpenAPI definition in the prompt prefix as described in Section~\ref{sec:prompt-engineering-methods}. The experiment was performed on the 7-billion-parameter Code Llama model with a 60/40 prefix-to-suffix ratio and a range of context sizes from 4\,096 to 7\,168 tokens. The extended prefix size was expected to compensate for the space taken by the components section in the prompt.

\begin{figure}[!h]
\centering
\includegraphics[width=\columnwidth]{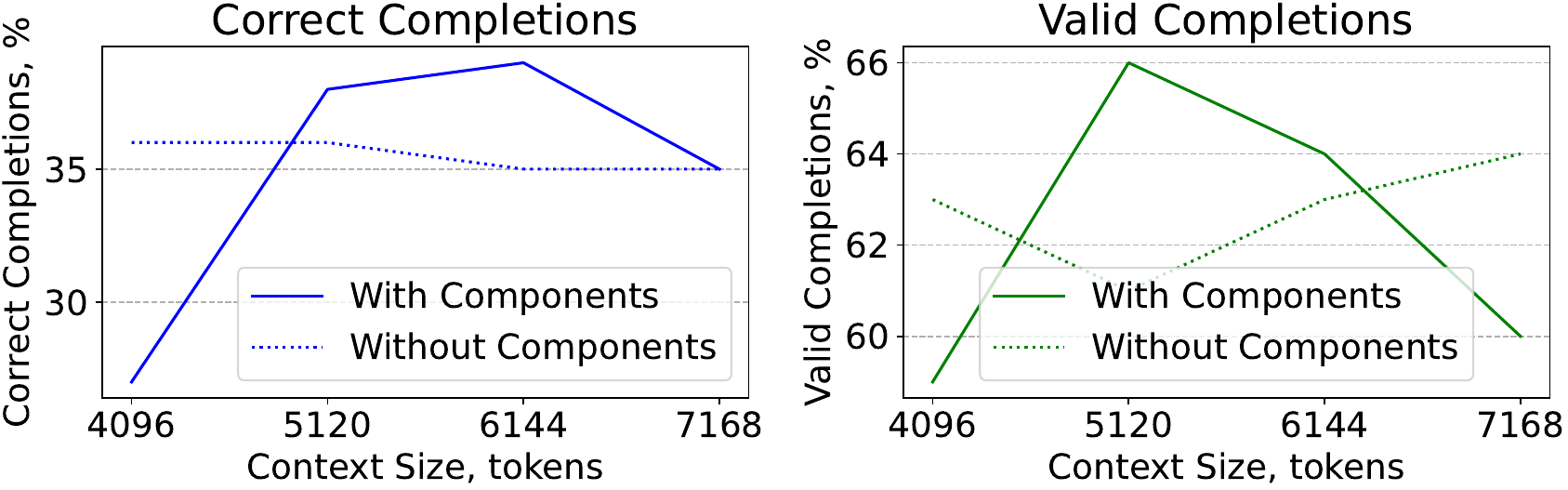}
\caption{OpenAPI completion performance of the original Code Llama 7B model with and without OpenAPI components in the prompt prefix.\label{fig:components-in-prefix}}
\end{figure}

As can be seen in Figure~\ref{fig:components-in-prefix}, adding the components list to the prompt prefix results in moderate validness and correctness improvements at the larger context sizes (11.4\% correctness and 1.6\% validness improvement at 6\,144 tokens). The results at a smaller context size of 4\,096 tokens, on the other hand, demonstrate a significant drop in both correctness (25\%) and validity (6.3\%).

The obtained results might mean that the components are not necessarily required for correct OpenAPI completion. The model can deduct the correct component names from just the surrounding context. When used with a relatively small context size, the suggested method can even be harmful to the solution performance as it consumes space that could otherwise be taken by more valuable information in the prefix.

\subsubsection{\bf Optimal model size}

The Code Llama model is available in four sizes: 7, 13, 34, and 70 billion parameters. The impact of the model size on the solution performance is evaluated by comparing the performance of the 7 and 13-billion-parameter models. Due to the resource constraints and the limitations of the larger models discussed in Section~\ref{sec:code-llama}, the 34 and 70-billion-parameter models were not evaluated.

A machine with four Nvidia A10G GPUs and a total of 96GB of GPU memory was used for the evaluation. This machine made it possible to evaluate the model with up to 4\,096 tokens of context. The evaluation was performed on a range of context sizes from 1\,024 to 4\,096 tokens with a constant prefix-to-suffix ratio of 50/50.

\begin{figure}[!h]
\centering
\includegraphics[width=\columnwidth]{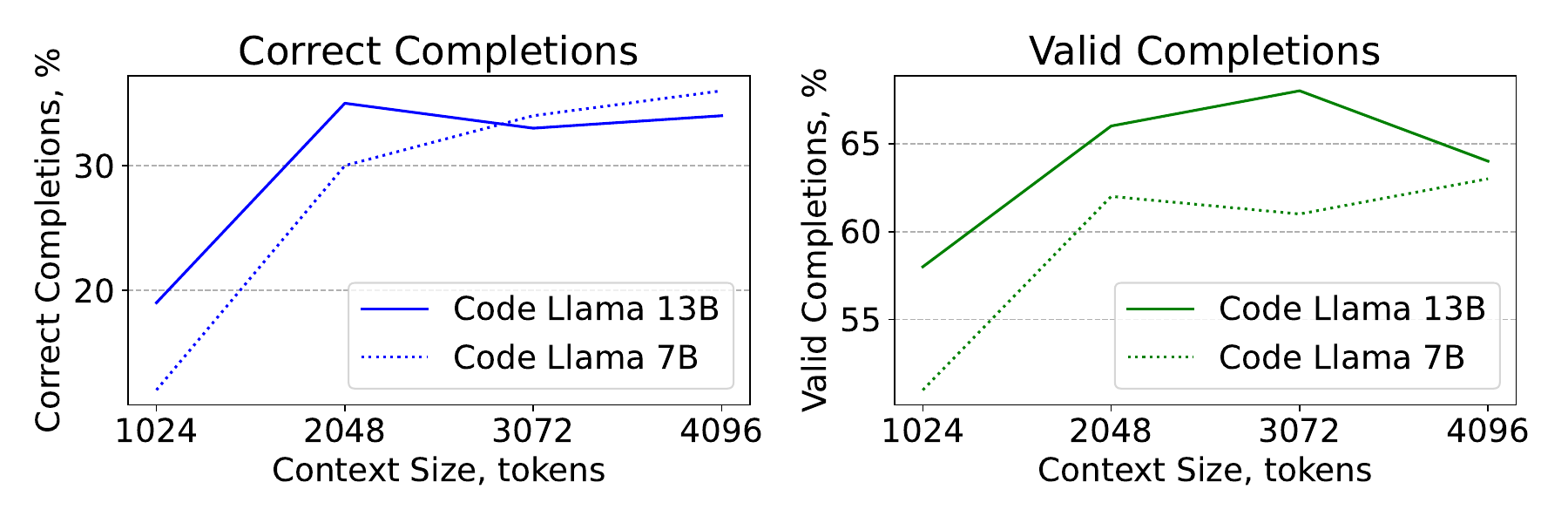}
\caption{OpenAPI completion performance of the 7 and 13-billion-parameter versions of the original Code Llama model.\label{fig:model-size}}
\end{figure}

As can be seen in Figure~\ref{fig:model-size}, the bigger 13-billion-parameter model outperforms the smaller model at all context sizes with the only exception being correctness after 3\,072 tokens. At the context size of 4\,096 tokens, Code Llama 13B surpasses the results obtained with Code Llama 7B by only 1\% in validity while falling behind in correctness by 5\%.

\begin{figure}[!h]
\centering
\includegraphics[width=\columnwidth]{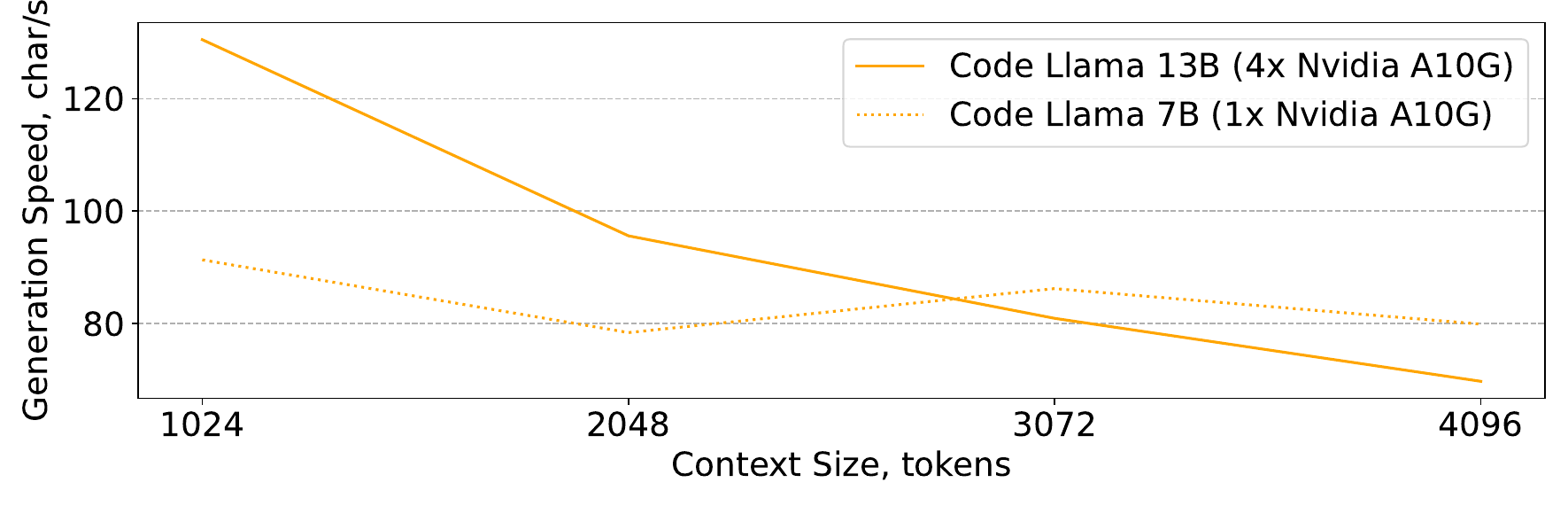}
\caption{Generation speed of the 7 and 13-billion-parameter versions of the original Code Llama model.\label{fig:model-size-speed}}
\end{figure}

Figure~\ref{fig:model-size-speed} compares the generation speed trends of the models. Due to the GPU computing power redundancy caused by the high memory requirements of the model, the 13 billion Code Llama outperforms the smaller model at 1\,024 and 2\,048 tokens. However, the quadratic complexity of the attention mechanism quickly makes up for the computing power redundancy, and the generation speed drops below the speed of the smaller model at 3\,072 and 4\,096 tokens. The speed reduction trend suggests that the difference in the generation speed will only increase with the context size.

\subsection{Fine-tuning Code Llama Model for OpenAPI Completion}

As demonstrated in the previous section, the original Code Llama model is already capable of OpenAPI completion and even outperforms GitHub Copilot. This section evaluates the fine-tuning approaches that can further improve the model performance.

\subsubsection{\bf Benefits of using a mixture of PSM and SPM infilling formats}

The study of Bavarian et al. suggests that using a mix of SPM and PSM formats during training leads to an increase in the infilling performance \cite{bavarian2022efficient}. At the same time, the original Code Llama model has a known problem with infilling in the SPM format \ref{sec:spm-vs-psm}. It is worth evaluating the efficiency of including this format in the fine-tuning process. This section compares the performance of the 7-billion-parameter Code Llama model fine-tuned solely with the PSM format and with a mixture of PSM and SPM formats (joint fine-tuning). When used, the SPM format was applied to half of the training samples. Context size of 4\,096 tokens and a 50/50 prefix-suffix split was used during fine-tuning.

\begin{figure}[!h]
\centering
\includegraphics[width=\columnwidth]{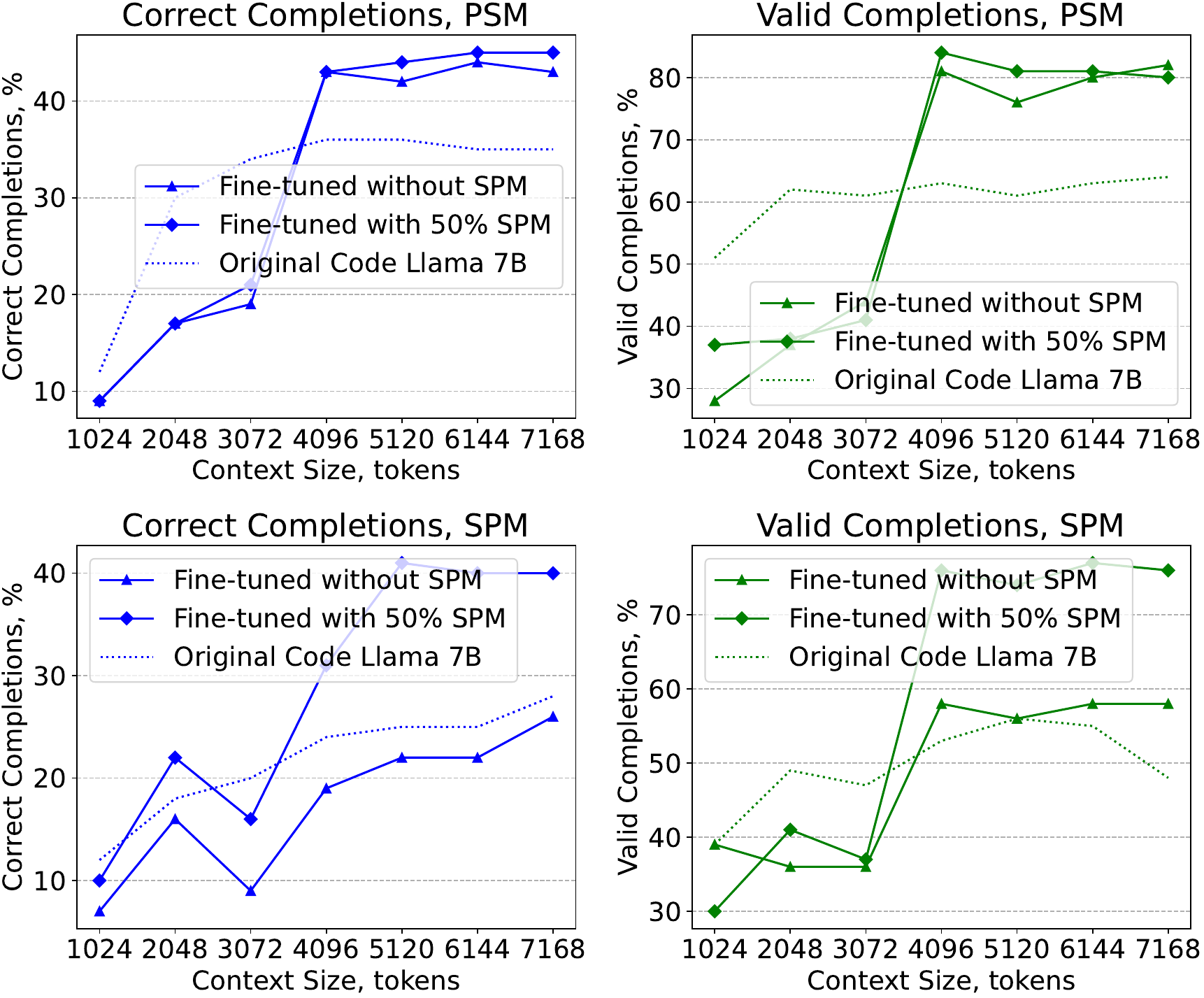}
\caption{OpenAPI completion performance of the Code Llama 7B model fine-tuned solely with the PSM format and with a mixture of PSM and SPM formats.\label{fig:fine-tuning-with-and-without-spm}}
\end{figure}

As can be seen in Figure~\ref{fig:fine-tuning-with-and-without-spm}, both fine-tuning approaches demonstrate similar performance in the PSM format with the joint fine-tuning showing a slight improvement in correctness and validity. The joint fine-tuning also results in significantly higher performance when the SPM format is used for inference. These results suggest that it always makes sense to include the SPM format when fine-tuning the Code Llama model for code infilling tasks.

Compared to the original Code Llama model, the model fine-tuned with the joint format demonstrates a peak improvement of 28.6\% in both correctness and validity at the context size of 6\,144 tokens when prompted with the PSM format. Performance in the SPM format is still inferior to PSM but the gap is reduced from 28.6\% to 11.1\% in correctness and from 12.7\% to 4.9\% in validity at the same context size. This shows that the selected fine-tuning approach can also lead to solving the original model's problem of split-token handling in the SPM format.

\subsubsection{\bf Training-time context size infilling bias and document splitting technique}

As revealed in the previous section, the fine-tuned model underperforms when prompted with context sizes smaller than used during fine-tuning (Figure~\ref{fig:fine-tuning-with-and-without-spm}). This can be caused by the bias towards training-time context size infilling discussed in Section~\ref{sec:fine-tuning-methods}. This section further evaluates the impact of the fine-tuning context size as well as the suggested document splitting technique on the performance of the fine-tuned model.

\begin{figure}[!h]
\centering
\includegraphics[width=\columnwidth]{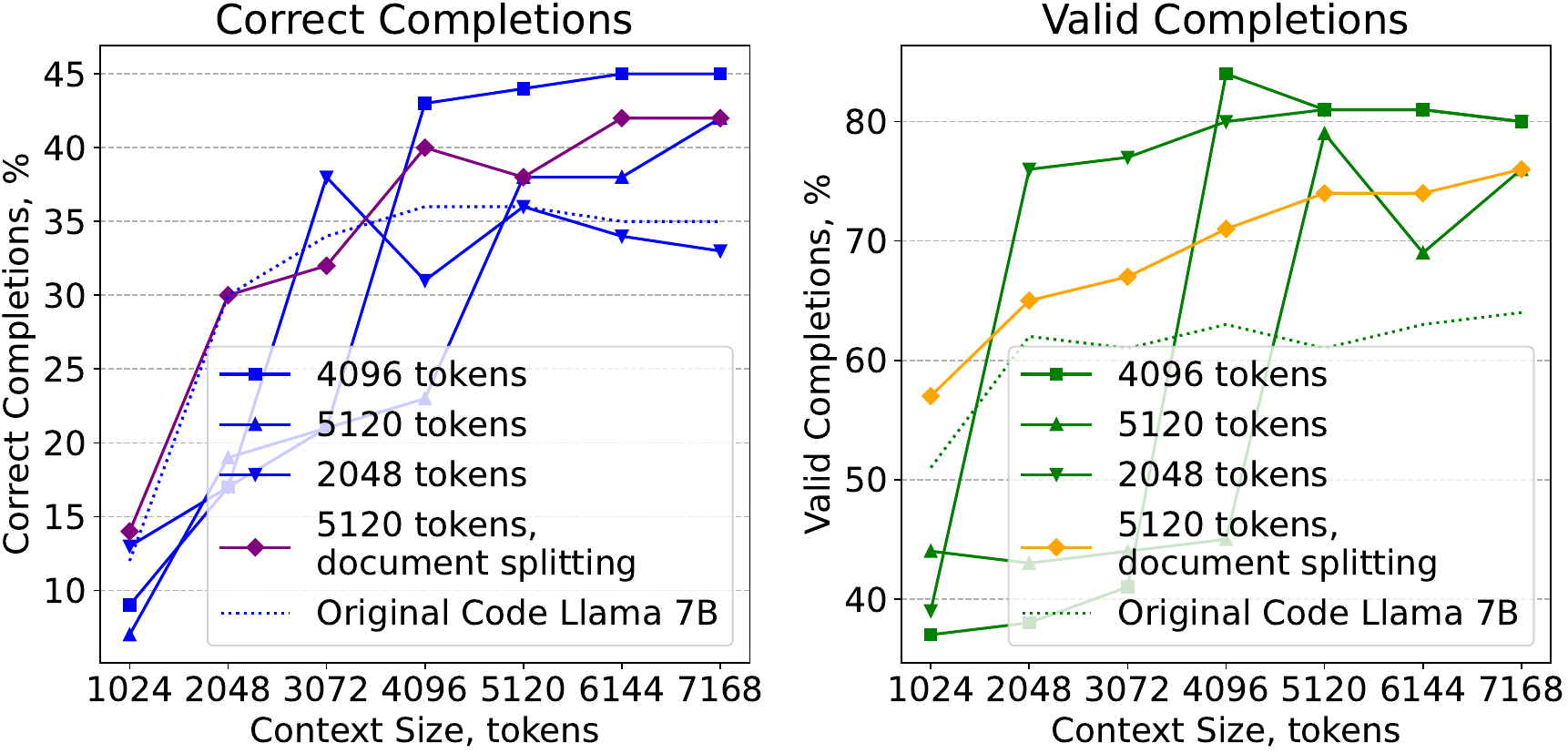}
\caption{OpenAPI completion performance of the Code Llama 7B model fine-tuned at different context sizes and with the document splitting technique.\label{fig:document-splitting}}
\end{figure}

As can be seen in Figure~\ref{fig:document-splitting}, the fine-tuned models with both 4\,096 tokens and 5\,120 tokens underperform compared to the original Code Llama model at context sizes smaller than the corresponding fine-tuning context sizes. After this point, the performance of the fine-tuned models stabilizes. Detailed inspection of the infilling results shows that the fine-tuned models tend to generate completions that are too long and contain irrelevant information when prompted with smaller context sizes.

A tempting idea to mitigate the problem is to use a much smaller context size for fine-tuning. As demonstrated in Figure~\ref{fig:document-splitting}, using the context size of 2\,048 tokens for fine-tuning indeed decreases the performance decline at smaller context sizes. However, using smaller context also reduces the performance gains at larger context sizes to the point where the fine-tuned model underperforms compared to the original Code Llama model. Nevertheless, the model still demonstrates the best validity rates at all context sizes.

The model fine-tuned with the document splitting technique demonstrates more equalized performance across the context sizes. The correctness and validity of the completions generated by this model surpass the original Code Llama model at almost all context sizes with a peak increase of 20\% at 6\,144 and 7\,168 tokens and a marginal decrease of 5.8\% at 3\,072 tokens. The model slightly underperforms compared to the model fine-tuned with the conventional context-level FIM technique at the context size of 4\,096 tokens when prompted with context sizes larger than 4\,096 tokens, but it outperforms the same model at smaller context sizes by a significant margin.

\section{Findings and Implications}

Through the course of this research, the impact of multiple variables, prompt engineering, and fine-tuning techniques on the OpenAPI completion performance of the Code Llama model was evaluated. The most notable experiment results are summarized and compared in Table~\ref{tab:performance-comparison}.

\subsubsection{\bf Original Code Llama model is capable of outperforming GitHub Copilot}

The original Code Llama 7B model, when used with the optimal prefix-to-suffix ratio of 50/50, the PSM prompt format, and the context size of 4\,096 tokens, outperforms GitHub Copilot by 24.1\% in correctness, generating correct OpenAPI completions 36\% of the time. At the same time, the model produces valid completions 63\% of the time which is a 7.3\% decrease compared to Copilot. It is worth pointing out that the GPT-3 model, which is the foundational model used by GitHub Copilot, uses 175 billion parameters which is 25 times more than in the case of the Code Llama 7B model.

\subsubsection{\bf Larger model does not always show significant performance improvement}

The larger 13-billion-parameter version of the Code Llama model demonstrated only a marginal correctness improvement compared to the smaller 7-billion-parameter version. This might mean that the OpenAPI format is not complex enough to leverage the additional model capacity. Considering the quadrupled computational cost, this renders the larger version of the model not practical for use in OpenAPI completion. 

\subsubsection{\bf Task-specific fine-tuning can significantly improve the model performance}

The 7-billion-parameter version of the Code Llama model fine-tuned for OpenAPI completion at the context size of 4\,096 tokens following the methods described in Section~\ref{sec:fine-tuning-methods} demonstrates a maximum correctness improvement of 28.6\% compared to the original model and further outperforms GitHub Copilot by 55.2\%. 

\subsubsection{\bf Conventional context-level FIM training objective can lead to suboptimal performance at smaller context sizes}

The case of OpenAPI completion revealed that, when used with documents that are significantly larger than the fine-tuning context size, this technique leads to a bias towards training-time context size infilling. This bias can cause the model to generate too long and irrelevant completions when prompted with a context size smaller than the fine-tuning context size. The models trained with this technique still demonstrate the best performance at the context size used during fine-tuning which can be leveraged when the inference context size is known in advance.

\subsubsection{\bf Document splitting technique leads to a more unified performance across context sizes}

The document splitting technique proposed in Section~\ref{sec:fine-tuning-methods} can be used to mitigate the problem of the training-time context size infilling bias and lead to a more unified performance across the context sizes. Even though the models fine-tuned without document splitting outperform the model fine-tuned with this technique at the context sizes used during fine-tuning, the latter model demonstrates the best average performance across the context sizes which is crucial for training foundational or open-source models that are expected to be used with a variety of context sizes.

\begin{table*}[!t]
\caption{OpenAPI Completion Performance Comparison\label{tab:performance-comparison}}
\centering
\begin{tabular}{lllrr}
\toprule
\textbf{Model / Solution} & \textbf{Correctness, max., \%} & \textbf{Validness, max., \%} & \textbf{Correctness, avg., \%} & \textbf{Validity, avg., \%} \\
\midrule
GitHub Copilot & 29 & 68 & 29.0 & 68.0 \\
Code Llama 7B & 36 (4\,096 tokens) & 64 (7\,168 tokens) & 31.1 & 60.7 \\
Code Llama 13B (evaluated at 1\,024-4\,096 tokens) & 34 (4\,096 tokens) & 68 (3\,072 tokens) & 30.2 & 64.0 \\
Code Llama 7B, fine-tuned at 4\,096 tokens & \textbf{45} (6\,144 tokens) & \textbf{84} (4\,096 tokens) & 32.0 & 63.1 \\
Code Llama 7B, fine-tuned at 5\,120 tokens & 42 (7\,168 tokens) & 79 (5\,120 tokens) & 26.8 & 57.1 \\
Code Llama 7B, fine-tuned with document splitting & 42 (6\,144 tokens) & 76 (7\,168 tokens) & \textbf{34.0} & \textbf{69.1} \\
\bottomrule
\end{tabular}
\end{table*}

\section{Conclusion}

In the course of this research, the first semantics-aware OpenAPI completion benchmark was proposed. The development of such a benchmark made it possible to evaluate the task-specific performance of one of the most widely used commercial code completion solutions -- GitHub Copilot.

A series of experiments performed in this research demonstrate how modern compact open-source models can be optimized to outperform commercial solutions. The state-of-the-art code generation model Code Llama from Meta is used here for that. 

The Code Llama model fine-tuned for OpenAPI completion reaches the peak correctness improvement of 55.2\% compared to GitHub Copilot. At the time of writing, this is the first work investigating the details of fine-tuning the Code Llama model for code infilling.

In the course of the experiments, an improvement to a widely used code infilling training technique was proposed. The optimization improves the stability of the fine-tuned model performance at the entire range of context sizes. This optimization is particularly beneficial for formats like OpenAPI, where the training document size is significantly larger than the fine-tuning context size.

The OpenAPI completion benchmark as well as the fine-tuning pipeline for the Code Llama model were made publicly available at \url{https://github.com/BohdanPetryshyn/openapi-completion-benchmark} and \url{https://github.com/BohdanPetryshyn/code-llama-fim-fine-tuning}. The authors hope that these resources will be found helpful by practitioners even outside of the realm of API design tooling.

\bibliographystyle{IEEEtran}
\bibliography{refs}

\newpage

\vfill

\end{document}